\documentclass[preprint,showpacs,preprintnumbers,amsmath,amssymb,prd]{revtex4}

\usepackage[dvips,final]{graphicx}
\usepackage{epsfig}
\usepackage{amsmath}

\begin{document}

\preprint{WUE-ITP-2002-020}
\preprint{MZ-TH-02-023}

\title{Nonfactorizable corrections to  $B \rightarrow J/\psi K$}

\author{Bla\v zenka Meli\'c}\thanks{
Alexander von Humboldt fellow. On the leave from the Rudjer Bo\v skovi\'c
Institute, Zagreb, Croatia}
\affiliation{%
Institut f\"ur Physik der Johannes-Gutenberg-Universit\"at,
\\
D--55099 Mainz, Germany\\
Institut f\"ur Physik und Astrophysik, Universit\"at W\"urzburg,
\\
D--97074 W\"urzburg, Germany
}%

\date{\today}% It is always \today, today,

\begin{abstract}
We apply the method of QCD light-cone sum rules to calculate nonfactorizable
contributions to the $B \rightarrow J/\psi K$ decay and
estimate soft nonfactorizable corrections to the $a_2$ parameter.  The corrections appear to be 
positive,
favoring the positive sign of $a_2$,
in agreement with recent theoretical considerations and experimental data. 
Our result also confirms expectations that in the color-suppressed decay nonfactorizable 
corrections are sizable.
\end{abstract}


\maketitle

%%%%%%%%%%%%%%%%%%%%%%%%%%%%%%%%%%%%%%%%%%%%%%%%%%%%%%%%%%%%%%%5                                
\newpage 

\section{\label{sec:intro}Introduction}

In nonleptonic decays of a $B$ meson one can study 
effects of
hadronization, perturbative as well as nonperturbative
dynamics, final state interaction
effects and CP violation. Measurements of exclusive nonleptonic $B$ decays have reached sufficient precision
to challenge our theoretical knowledge on such decays. It became clear that
calculations have to reduce their theoretical 
uncertainties in order to make real use of data.
Nowadays there exist several approaches 
which shed more light on the dynamical background of exclusive nonleptonic decays. The
most exploited ones are {\it QCD factorization} \cite{BBNS} 
and {\it PQCD approach} \cite{KLS}. 
The PQCD model assumes that the two-body nonleptonic
amplitude is perturbatively calculable if the Sudakov suppression is implemented to the calculation.
In QCD factorization one can  
 show the factorization of the
weak decay amplitude at the leading order $1/m_b$ level and 
can consider systematically perturbatively calculable
nonleading terms of $1/m_b$ expansion.
None of these approaches can take nonperturbative ${\cal O}(1/m_b)$ terms into account, but there is
no evidence that such terms are negligible. 

The $B \rightarrow J/\psi K$ decay is interesting because of the several reasons.
There is a large discrepancy between the experiment
and {\it the (naive) factorization} prediction. 
The naive factorization is based on the assumption that the nonleptonic amplitude (obtained 
in terms of matrix elements of four-quark
operators by using the effective weak Hamiltonian) can be
expressed as a product of matrix elements of two hadronic (bilinear) currents.
It also predicts
vanishing matrix elements of four-quark operators with the mismatch of the color indices.
The naive factorization hypothesis has been confirmed experimentally only for
class-I $B \to D^{(*)}  M$ $(M=\pi,\rho, a_1, D_s, D_s^*)$ decays. 
On the other hand, $B \rightarrow J/\psi K$ is
the color-suppressed (class-II) decay and therefore a significant impact of nonfactorizable
contributions is expected.

Effects of a violation of the factorization hypothesis in the $B \rightarrow J/\psi K$ mode
have been, up to now,
calculated by using different theoretical methods, resulting in the sign ambiguity of the
decay amplitude i.e. $a_2$ parameter ($a_2$ parameter 
is the effective coefficient of four-quark
operators in the weak Hamiltonian;
it is defined below by eqs. (\ref{eq:result}) and (\ref{eq:a2def})). 
The QCD sum rule approach \cite{KR2} 
predicted a negative value for $a_2$, while the
PQCD hard scattering approach \cite{LY} and the calculation done
in QCD factorization \cite{Cheng} gave the positive value for the
$a_2$ parameter. 
Moreover, a detailed analysis of the experimentally
determined $B$ meson branching ratios, although by assuming the universality of the 
$a_2$ parameter, gives conclusive evidence that
generally the $a_2$ parameter should be positive \cite{NS}. 
On the contrary, the negative value of $a_2$ would indicate that
the $1/N_c$ term and the nonfactorizable part in the amplitude tend to cancel and would
therefore confirm the large $N_c$ hypothesis \cite{BGR}. The validity of this hypothesis was 
established in two-body $D$ meson decays, while,  
up to now, different attempts failed to
prove this assumption for $B$ decays (see i.e discussion in \cite{KR1}). 
However, the sign ambiguity of $a_2$ cannot be solved experimentally by
considering the $B \rightarrow J/\psi K$ decay alone. One of the possibilities is 
to consider the interference between the short- and long-distance contributions to
$B \rightarrow K l^+l^-$  \cite{Soares}. 

Nonfactorizable corrections due to the exchange of hard gluons were calculated at 
$O(\alpha_s)$ in QCD factorization. In this paper we concentrate
on the LCSR 
estimation of soft nonfactorizable contributions in the $B \rightarrow J/\psi K$ 
decay coming from the
exchange of soft gluons between the $J/\psi$ and the kaon. 
The calculation is based on  {\it the light-cone sum rules (LCSR) method} 
\cite{Khodja}. This method enables a consistent calculation of 
nonperturbative corrections of hadronic amplitudes inside the same
framework reducing therefore the model uncertainties.

The paper is structured as follows. 
First, we discuss the results of the (naive) factorization method in Section~\ref{sec:fac}. 
The LCSR for the
$B \rightarrow J/\psi K$ decay is derived in Section~\ref{sec:sum}. 
Next, to show the consistency of the method,
we prove the factorization of the leading order contribution in Section~\ref{sec:fact}. 
In Section~\ref{sec:calc}, the calculation of the
soft nonfactorizable corrections is done by including twist-3 and twist-4 contributions. 
Section~\ref{sec:num} assembles the numerical results. 
In Section~\ref{sec:comp}, using the
results of calculation, we discuss impact of the nonfactorizable term 
on the factorization assumption and
the implications of the results. A conclusion is given in Section~\ref{sec:conc}.

\section{\label{sec:fac}Factorization hypothesis and nonfactorizable contributions}

The part of the effective weak Hamiltonian relevant for the 
$B \rightarrow J/\psi K$ decay can be
written in the form 
\begin{equation}
H_W = \frac{G_F}{\sqrt{2}} V_{c b} V_{c s}^*
\left[ C_1(\mu) {\cal O}_1 +  C_2(\mu) {\cal O}_2 \right] \, , 
\label{eq:ham0}
\end{equation}
which can be further expressed as
\begin{equation}
H_W = \frac{G_F}{\sqrt{2}} V_{c b} V_{c s}^*
\left[ \left ( C_2(\mu) + \frac{C_1(\mu)}{3} \right ) {\cal O}_2 +  2 C_1(\mu) {\cal \tilde{O}}_2 \right] \, 
\label{eq:ham}
\end{equation}
where
\begin{equation}
{\cal O}_2 = (\overline{c} \Gamma_{\mu} c)(\overline{s} \Gamma^{\mu} b) \;\; ,
{\cal \tilde{O}}_2 =  (\overline{c} \Gamma_{\mu} \frac{\lambda_a}{2} c)
(\overline{s} \Gamma^{\mu} \frac{\lambda_a}{2} b) \, . 
\label{eq:oper}
\end{equation}
Here $\Gamma_{\mu} = \gamma_{\mu} (1-\gamma_5)$, $V_{ij}$'s are CKM matrix elements and 
$\lambda^a$'s are $SU(3)$ color matrices. $C_1(\mu)$ and $C_2(\mu)$ are short-distance 
Wilson coefficients computed at the renormalization scale $\mu \sim O(m_b)$. 
The ${\cal \tilde{O}}_2$ operator appears after the projection of the color-mismatched quark fields in
${\cal {O}}_1 = (\overline{c} \Gamma_{\mu} b)(\overline{s} \Gamma^{\mu} c) $ to a color singlet state:
\begin{equation}
{\cal {O}}_1 = \frac{1}{N_c} {\cal {O}}_2 + 2 {\cal \tilde{O}}_2 \, . 
\end{equation}
The $1/N_c$ term is the origin of the factor $1/3$ in (\ref{eq:ham}).

Under the assumption that the matrix element for the $B \rightarrow J/\psi K$ decay 
factorizes, the matrix element of the $ {\cal \tilde{O}}_2$ operator vanishes
because of the color conservation and the rest can written as 
\begin{eqnarray}
\langle J/\psi K | H_W | B \rangle &=& \frac{G_F}{\sqrt{2}} V_{c b} V_{c s}^*
\left [C_2(\mu) + \frac{C_1(\mu)}{3} \right ] \nonumber \\
& & \hspace*{-2cm} \times 
\langle  J/\psi K |  {\cal O}_2 | B \rangle^{fact} \left [ 1 + O(\alpha_s) +
O \left (\frac{\Lambda_{QCD}}{m_b} \right ) \right ] \, , 
\label{eq:decay}
\end{eqnarray}
where the second and the third term represent hard and soft corrections to the 
factorizable amplitude, respectively. 

The factorized matrix element of the operator ${\cal O}_2$ is given by 
\begin{eqnarray}
\langle  J/\psi(p) K(q) |  {\cal O}_2 | B(p+q) \rangle^{fact}
&=& \langle J/\psi(p) | \overline{c} \Gamma_{\mu} c
| 0 \rangle
\langle K(q)| \overline{s} \Gamma^{\mu} b | B(p+q) \rangle \nonumber \\
&=& 2 \epsilon \cdot q \, m_{J/\psi}  f_{J/\psi} F^+_{BK}(m_{J/\psi}^2) \, 
\label{eq:fac}
\end{eqnarray}
where meson momenta are explicitly specified and $p^2 = m_{J/\psi}^2$. 
The $J/\psi$ decay constant is defined by the relation
\begin{equation}
\langle J/\psi(p) | \overline{c} \gamma_{\mu} c | 0 \rangle
= f_{J/\psi} m_{J/\psi} \, \epsilon_{\mu}
\end{equation}
with $\epsilon_{\mu}$ being the $J/\psi$ polarization vector which satisfies the condition
$\epsilon \cdot p = 0$. 
The $ f_{J/\psi}$ denotes the $J/\psi$ decay constant determined by the experimental leptonic width $\Gamma (J/\psi
\rightarrow l^+l^-) = 5.26 \pm 0.37 \; {\rm keV}$ by using the leading order calculation: 
\begin{equation}
f_{J/\psi} = 405 \pm 14 \, {\rm MeV}\, .
\label{eq:fpsi}
\end{equation}
The $F_{BK}^+$ form factor is defined through the decomposition
\begin{equation}
\langle K(q)| \overline{s} \gamma_{\mu} b | B(p+q) \rangle =  (2 q + p)_{\mu}  F^+_{BK}(p^2) +
\frac{ m_B^2 - m_K^2 }{q^2} p_{\mu} (- F^+_{BK}(p^2) + F^0_{BK}(p^2) )
\end{equation}
and estimated from the light-cone sum rules \cite{BKR,KRWWY} has the value
\begin{equation}
F^+_{BK}(m_{J/\psi}^2) = 0.55 \pm 0.05\, .  
\label{eq:f0}
\end{equation}

By neglecting corrections in (\ref{eq:decay}), the (naive) factorization 
expression for the $B \rightarrow J/\psi K$ decay emerges. 
Taking into account the NLO Wilson coefficients calculated in the
NDR scheme \cite{BBL}
for $\mu = \overline{m_b}(m_b) = 4.40 {\rm \,
GeV}$ and $\Lambda_{\overline{\rm MS}}^{(5)} = 225\, {\rm MeV}$, 
\begin{eqnarray}
C_1(\overline{m_b}(m_b)) = 1.082 \,\qquad C_2(\overline{m_b}(m_b)) = -0.185
\label{eq:cc}
\end{eqnarray}
and using the $B$ meson lifetime $\tau(B^{\pm}) = 1.653 \pm 0.28 \,  ps$, we obtain for the 
branching ratio in
the naive factorization 
\begin{equation}
{\cal B}(B \rightarrow J/\psi K)^{fact} = 3.3 \cdot 10^{-4}\,,
\label{eq:BRnf}
\end{equation}
with the uncertainties in the order of $30\%$. 
This has to be compared with the recent measurements \cite{exp}
\begin{eqnarray}
{\cal B}(B^+ \rightarrow J/\psi K^+) &=& (10.1 \pm 0.3 \pm 0.5) \cdot 10^{-4}\, , 
\nonumber \\
{\cal B}(B^0 \rightarrow J/\psi K^0) &=& (8.3 \pm 0.4 \pm 0.5)\cdot 10^{-4} \, . 
\label{eq:BRexp}
\end{eqnarray}
Obviously there is a large discrepancy between the naive factorization prediction ({\ref{eq:BRnf}}) and 
the experiment. 

Returning to the expression for the $B \rightarrow J/\psi K$ amplitude (\ref{eq:decay}), 
the corrections are given as an expansion in $1/m_b$ and $\alpha_s$.  
Apart from the $O(\alpha_s)$ corrections to the factorizable part, 
there are also nonfactorizable corrections, which can be, either due to a hard gluon 
exchange 
or due to a soft gluon exchange 
(denoted in (\ref{eq:decay}) as O($\alpha_s$) or $O(\Lambda_{QCD}/m_b)$ corrections, 
respectively) 
between $J/\psi$ and the $B-K$ system. 

To be able to discuss the impact of the nonfactorizable terms, it is usual to
parameterize the $\langle J/\psi K | H_W | B \rangle$ amplitude in terms of the $a_2$ parameter as 
\cite{NS,KR1} 
\begin{equation}
\langle J/\psi K | H_W | B \rangle = \sqrt{2}\, G_F \, V_{c b} V_{c s}^* \, \epsilon \cdot q \,
m_{J/\psi} f_{J/\psi} F_{BK}^+(m_{J/\psi}^2)\, a_2 \, . 
\label{eq:result}
\end{equation}
The effective parameter $a_2$ is defined by
\begin{equation}
a_2 = C_2(\mu) + \frac{C_1(\mu)}{3} +
2 C_1(\mu) 
\frac{\tilde{F}_{BK}^+(\mu)}{F_{BK}^+(m_{J/\psi}^2)}  \, . 
\label{eq:a2def}
\end{equation}
%$F(\mu)^{hard}$ denotes hard nonfactorizable contributions and 
The part proportional to the $\tilde{F}_{BK}^+$ represents the nonfactorizable contribution from
the ${\cal \tilde{O}}_2$ operator
\begin{equation}
\langle J/\psi K | {\cal \tilde{O}}_2(\mu) | B \rangle = 2 \epsilon \cdot q \,
m_{J/\psi} f_{J/\psi} \tilde{F}_{BK}^+(\mu^2)
\end{equation}
and  $\tilde{F}_{BK}^+ =0$ corresponds to the naive factorization result,
Eq. (\ref{eq:fac}). 

Because there is no 
explicit $\mu$ dependence of matrix element (\ref{eq:fac}), the $\mu$ dependence of 
$a_2$ needs to be canceled by the nonfactorizable term. 
The nonvanishing nonfactorizable part is 
also required in order to suppress the strong renormalization 
scheme dependence of the effective parameter 
$a_2$ \cite{Buras}. 

Using the parameterization (\ref{eq:result}) we can extract the $a_2$ coefficient from the experiment.  
From the measurements (\ref{eq:BRexp}) one obtains
\begin{equation}
|a_2^{exp}| = 0.29 \pm 0.03 \, , 
\label{eq:a2exp}
\end{equation}
with the undetermined sign of $a_2$. 

On the other hand, with the NLO Wilson coefficients from (\ref{eq:cc}), 
the naive factorization yields 
\begin{equation}
a_{2, \, NLO}^{fact} = C_2(m_b) + \frac{C_1(m_b)}{3} = 0.176 \, , 
\label{eq:a2nf}
\end{equation}
which is significantly below the value extracted from the experiment. 

Following \cite{KR1}, in Fig.1 we show the partial width for $B \rightarrow J/\psi K$ as a function 
of the nonfactorizable amplitude $\tilde{F}_{BK}$. 
\begin{figure}
\begin{center}
\includegraphics[width=0.45\textwidth]{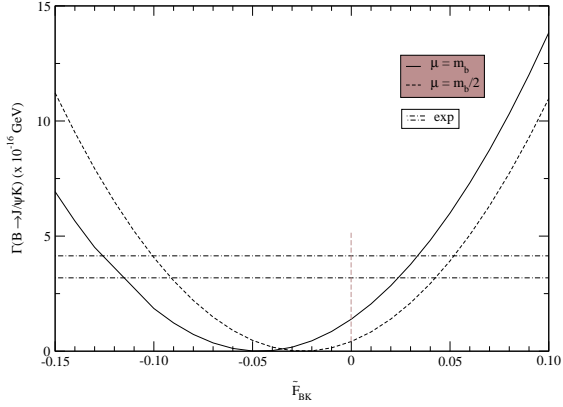}
\end{center}
\caption{The partial width $\Gamma (B \rightarrow J/\psi K)$ as a function 
of the nonfactorizable amplitude $\tilde{F}_{BK}$.}
\end{figure}
The zero value of $\tilde{F}_{BK}$ corresponds to the factorizable prediction. 
There exist
two ways to satisfy the experimental demands on the $\tilde{F}_{BK}$. According to the 
large $1/N_c$ rule assumption \cite{BGR}, one can argue that there is a cancellation between 
$1/N_c$ piece of the factorizable part and the nonfactorizable contribution, 
(\ref{eq:a2def}). That would demand relatively small and negative value of $\tilde{F}_{BK}$. The
other possibility is to have even smaller, but positive values for $\tilde{F}_{BK}$, which then 
compensate the overall smallness of the factorizable part and bring the theoretical 
estimation of $a_2$ in agreement with experiment. 

One can note significant        
$\mu$ dependence of the theoretical expectation for the partial width in Fig.1, which 
brings an uncertainty in the prediction for $\tilde{F}_{BK}(\mu)$ 
of the order of  $30 \%$. This uncertainty is
even more pronounced for the positive solutions of $\tilde{F}_{BK}(\mu)$.
The values for $\tilde{F}_{BK}^+$ extracted from experiments 
\begin{eqnarray}
\tilde{F}_{BK}^+(m_b) &=& 0.028 \qquad {\rm or} \qquad \tilde{F}_{BK}^+(m_b)= -0.120 \, , 
\nonumber \\
%\end{eqnarray}
%and
%\begin{eqnarray}
\tilde{F}_{BK}^+(m_b/2) &=& 0.046 \qquad {\rm or} \qquad \tilde{F}_{BK}^+(m_b/2)= -0.095 \, . 
\label{eq:fexp2}
\end{eqnarray}
%which gives a two-fold ambiguity of the result, reflecting the unknown sign of the $a_2$.
clearly illustrate the $\mu$ sensitivity of the nonfactorizable part. 

In the following we will calculate the nonfactorizable contribution $\tilde{F}_{BK}^+$,  
which appears due to the exchange of soft gluons,  by using the 
QCD light-cone sum rule method. 

\section{\label{sec:sum}Light-cone sum rule for $\langle J/\psi K| {\cal O} | B \rangle $ }

\subsection{The correlator}

To estimate the soft-gluon exchange contributions to $B \rightarrow J/\psi K$ we use the method developed in 
\cite{Khodja} for the $B \rightarrow \pi \pi$ case. In this approach one considers the 
correlation function:
\begin{equation}
F_{\nu}(p,q,k) = i^2 \, \int d^4 x e^{-i(p+q)x} \int d^4 y e^{i(p-k)y} \langle K(q) | 
T \{ j_{\nu}^{J/\psi}(y) {\cal O}(0) j_5^{B}(x) \} | 0 \rangle \, , 
\label{eq:corr0}
\end{equation}
where $j_{\nu}^{J/\psi} = \overline{c} \gamma_{\nu} c$ and $j_5^{B} = m_b \overline{b} i \gamma_5 u$ are 
currents interpolating the $J/\psi$ and $B^{-}$ meson fields, respectively. 
\begin{figure*}
\begin{center}
\includegraphics*[width=17cm]{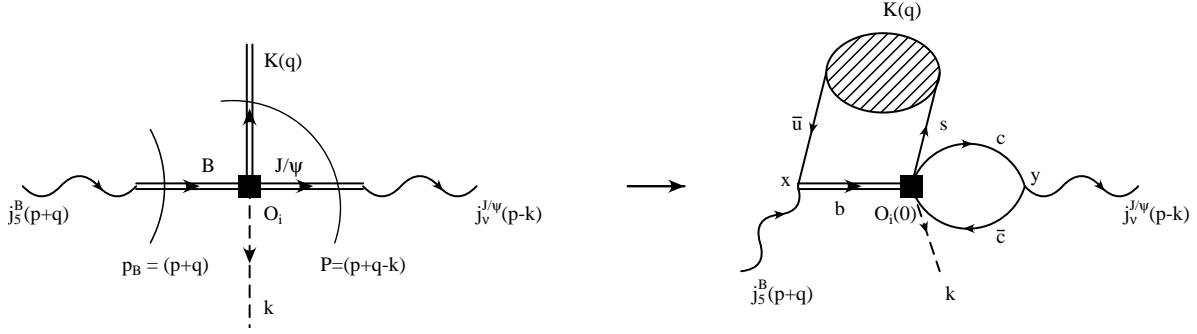}
\end{center}
\caption{$B \rightarrow J/\psi K$ decay in LCSR. The shaded oval region denotes nonperturbative 
input, the $K$ meson light-cone distribution amplitude. $J/\psi$ and the $B$ meson are represented 
by the currents $j^{J/\psi}(p-k)$ and $j^B(p+q)$, respectively. The square stands for the ${\cal O}_i$ four-quark 
weak operators.}
\end{figure*}
The correlator is a function of three independent momenta, 
chosen by convenience to be $q$, $p-k$ and $k$. 
Diagrammatically the correlator is shown in Fig.2. 

Here, it is important to emphasize the role of the unphysical $k$ momentum in the weak vertex. 
It was introduced in order to avoid that the $B$ meson four-momenta before ($p_B$) and 
after the decay ($P$) are the same, Fig.2.  
In such a way, one avoids a continuum of light contributions in the 
dispersion relation in the $B$-channel. These contributions, like 
$D\overline{D}^*_s$ or
$D^* \overline{D}_s$, 
have masses much smaller 
than the ground state $B$ meson mass and spoil the extraction of the physical $B$ state. 
Also, they are not exponentially suppressed by the Borel transformation 
(see for example the discussion in \cite{BS}). 

The correlator (\ref{eq:corr0}) for nonvanishing $k$ is a function of 
6 independent kinematical invariants. Four of them are taken to be the external momenta 
squared: $(p+q)^2$, $(p-k)^2$, $q^2$ and $k^2$, and additionally we take $P^2 = (p-k+q)^2$ and $p^2$. 
We neglect the small corrections of the order $O(m_K^2/m_B^2)$ and take $q^2 = m_K^2 =0$. 
Also, we set 
$k^2 =0$. 
The $p^2$ momentum is for the moment kept undefined, in order to be able to make unrestricted 
derivation of the sum rules. Its value is going to be set later, by considering 
the twist-2 calculation of the factorizable part, Section~\ref{sec:fact}, and will be chosen 
$p^2 = m_{J/\psi}^2$ in order to reproduce the factorization result, (\ref{eq:fac}).
Furthermore, we take $(p-k)^2$, $(p+q)^2$ and $P^2$ spacelike and 
large in order to stay away 
from the hadronic thresholds in both, the $J/\psi$ and the $B$ channel. All together we have
\begin{equation}
q^2 = k^2 = 0, \; p^2 = m_{J/\psi}^2, \; |(p-k)|^2 \gg \Lambda_{QCD}, |(p+q)|^2 \gg\Lambda_{QCD},  |P|^2 \gg \Lambda_{QCD}
\, . 
\end{equation}

\subsection{Derivation}

The first step is the derivation of the dispersion relation from the correlator (\ref{eq:corr0}).  
Inserting a complete 
set of hadronic states with the $J/\psi$ quantum numbers between the $J/\psi$ current and the 
weak operator in (\ref{eq:corr0})  gives us the following:
\begin{eqnarray}
F_{\nu} &=& 
\frac{m_{J/\psi} f_{J/\psi}}{ m_{J/\psi}^2 - (p-k)^2} \sum_{\lambda} \epsilon_{\nu}^{\ast (\lambda)}
\Pi( (p+q)^2, P^2, p^2, \epsilon^{(\lambda)}\cdot q_i) \nonumber \\
&+& \int_{s_0^{h(J/\psi)}}^{\infty} ds 
\frac{\rho_{h\, , \nu}^{J/\psi}(s,(p+q)^2,P^2,p^2)}
{s - (p-k)^2}  \, , 
\label{eq:hadr1}
\end{eqnarray}
where $q_i = p,q,k$ and $\epsilon^{(\lambda)}\cdot (p-k)=0$. The sum runs over the polarizations of $J/\psi$. 
The lowest state contribution satisfies
\begin{equation}
\langle 0 |\overline{c}\gamma_{\nu} c | J/\psi (p-k,\epsilon^{(\lambda)}) \rangle = m_{J/\psi} f_{J/\psi} 
\epsilon^{\ast (\lambda)}_{\nu}
\end{equation}
and $(p-k)^2 = m_{J/\psi}^2$. 
In (\ref{eq:hadr1}), $\rho_{h \, , \nu}^{J/\psi}$ and $s_0^{h(J/\psi)}$ are the spectral density and 
the threshold mass squared of the lowest excited resonances and continuum states of the $J/\psi$ channel, respectively. 

\begin{widetext}

The hadronic matrix element of interest is denoted by 
\begin{equation}
\Pi( (p+q)^2, P^2, p^2, \epsilon^{(\lambda)}\cdot q_i) = 
i \int d^4 x e^{-i(p+q)x} \langle J/\psi (p-k, \epsilon^{(\lambda)}) K(q) |
T \{ {\cal O}(0) j_5^{B}(x) \}|0 \rangle \, . 
\label{eq:pi}
\end{equation} 

On the other hand, for spacelike $(p-k)^2 \ll m_{J/\psi}^2$ far away from the poles 
associated with the resonances and continuum states, the correlator 
$F_{\nu}$ 
can be calculated in QCD in terms of the quark and gluon degrees of 
freedom and written in a form of a dispersion relation as:
\begin{equation}
{\cal F}_{\nu} = \frac{1}{\pi}
\int_{4 m_c^2}^{\infty} ds \frac{{\rm Im}_s {\cal F}_{\nu}
(s,(p+q)^2,P^2,p^2)}{s - (p-k)^2} \, ,  
\label{eq:f}
\end{equation}
with the kinematical decomposition 
\begin{equation}
{\cal F}_{\nu} = (p-k)_{\nu} F^{(p-k)} + k_{\nu} F^{(k)} + q_{\nu} F^{(q)} + 
\epsilon_{\nu \alpha \beta \gamma} (p-k)^{\alpha} k^{\beta} q^{\gamma}  F^{(\epsilon)} \, . 
\label{eq:fdec}
\end{equation}

By assuming quark-hadron 
duality one substitutes the hadronic spectral density $\rho_{h,\nu}^{J/\psi}$ in (\ref{eq:hadr1}) 
with the one 
calculable in QCD and replaces $s_0^{h{(J/\psi)}}$ with the effective 
threshold of the perturbative continuum, $s_0^{J/\psi}$, i.e :
\begin{equation}
\rho_{h\, , \nu}^{J/\psi}(s,(p+q)^2,P^2,p^2)\Theta(s-s_0^{h{(J/\psi)}}) =
\frac{1}{\pi} {\rm Im}_s {\cal F}_{\nu}(s,(p+q)^2,P^2,p^2)\Theta(s - s_0^{J/\psi}) \, . 
\label{eq:dpsi}
\end{equation}

By matching the hadronic relation (\ref{eq:hadr1}) with the QCD calculation (\ref{eq:f}) 
one obtains the sum rule expression
\begin{eqnarray}
 \frac{m_{J/\psi} f_{J/\psi}}{ m_{J/\psi}^2 - (p-k)^2} \sum_{\lambda} 
\epsilon_{\nu}^{\ast 
(\lambda)} \Pi((p+q)^2, P^2, p^2, 
 \epsilon^{(\lambda)}\cdot q_i) 
&=& \nonumber \\
& & \hspace*{-5cm} \frac{1}{\pi}\int_{4 m_c^2}^{s_0^{J/\psi}} ds 
\frac{{\rm Im}_s {\cal F}_{\nu}(s,(p+q)^2,P^2, p^2)}{s - (p-k)^2} \, . 
\label{eq:hadr2}
\end{eqnarray}

In order to reduce the impact of the approximation (\ref{eq:dpsi}) and to suppress contributions 
from excited and continuum states, as usually done for quarkonium systems 
one performs $n$ derivations in the momentum $(p-k)^2$ and 
receives $n$-moment sum rule for the correlator $\Pi((p+q)^2, P^2, p^2, \epsilon^{(\lambda)}\cdot q_i)$ of the form
\begin{eqnarray}
\sum_{\lambda} \epsilon_{\nu}^{\ast (\lambda)} \Pi((p+q)^2, P^2, p^2, \epsilon^{(\lambda)}\cdot q_i) 
&=& \nonumber \\
& & \hspace*{-5cm}  
\frac{1}{\pi \, m_{J/\psi} f_{J/\psi}}
\int_{4 m_c^2}^{s_0^{J/\psi}} ds \frac{(m_{J/\psi}^2 + Q_0^2)^{n+1}}{(s + Q_0^2)^{n+1}} 
{\rm Im}_s {\cal F}_{\nu}(s,(p+q)^2,P^2, p^2) \, , 
\label{eq:pibor}
\end{eqnarray}
where $Q_0$ is the sum rule parameter that role will be discussed later, in Section~\ref{sec:num}. 

We proceed by 
using the analytical properties of the $\Pi((p+q)^2, P^2, p^2,  
 \epsilon^{(\lambda)} \cdot q_i)$ amplitude in the $(p+q)^2$ variable of the $B$-channel 
and insert in (\ref{eq:pi}) the complete set of 
 hadronic states with the $B$ meson quantum numbers which yields 
\begin{eqnarray}
\sum_{\lambda} \epsilon_{\nu}^{\ast (\lambda)} \Pi((p+q)^2, P^2, p^2, 
 \epsilon^{(\lambda)}\cdot q_i ) &=& \nonumber \\
& &  \hspace*{-8cm} \frac{m_{B}^2 f_{B}}{m_{B}^2 - (p+q)^2} \sum_{\lambda} \epsilon_{\nu}^{\ast (\lambda)} 
\langle J/\psi(p-k, \epsilon^{(\lambda)}) K(q) | {\cal O}(0) | B(p+q) \rangle 
 + \int_{s_0^{h(B)}}^{\infty} ds' \frac{\rho_{h,\nu}^{B}(s',P^2,p^2)}{s' - (p+q)^2} \, . 
\nonumber \\
 \label{eq:piB}
\end{eqnarray}
In above, as before, it is assumed that in the last term the polarization sum is already done. 

The QCD part, given by the r.h.s of the eq.(\ref{eq:pibor}) and 
rewritten in a form of the dispersion relation, now 
in the $(p+q)^2$ variable, exposes the form of the double dispersion relation as
\begin{eqnarray}
\frac{1}{\pi}\int_{4 m_c^2}^{s_0^{J/\psi}} ds \frac{(m_{J/\psi}^2 + Q_0^2)^{n+1}}{(s + Q_0^2)^{n+1}} 
{\rm Im}_s {\cal F}_{\nu}(s,(p+q)^2,P^2,p^2) &=&  \nonumber \\
& & \hspace*{-10cm} \frac{1}{\pi^2}\int_{4 m_c^2}^{s_0^{J/\psi}} 
ds \frac{(m_{J/\psi}^2 + Q_0^2)^{n+1}}{(s + Q_0^2)^{n+1}} \int_{m_b^2}^{f_1(s,P^2,p^2)}  \frac{ds'}{s' - (p+q)^2}  
{\rm Im}_{s'} {\rm Im}_s {\cal F}_{\nu}(s,s',P^2,p^2)
\label{eq:dd}
\end{eqnarray}  
From the Maldestam representation of the kinetic variables one can see that the integration limit of $s'$ variable is going 
in general to depend on $s$, $P^2$ and $p^2$ and we denoted these dependence by 
$f_1(s,P^2,p^2)$. 
In the following, those terms which disappear after 
taking moments in $J/\psi$ 
channel and after making the Borel transform in the $B$-channel are neglected. 

In order to subtract the continuum of $B$ states, 
we exchange the order of the integration in (\ref{eq:dd}) and 
use quark-hadron
duality in $B$ channel in a sense that the spectral density $\rho_h^B$ is 
approximated by the $s' \ge s_0^B$ part of the double dispersion integral
(\ref{eq:dd}), 
where  $s_0^B$ is the effective 
threshold of the perturbative continuum in the $B$ channel. Therefore, 
\begin{eqnarray}
\frac{m_{B}^2 f_{B}}{ m_{B}^2 - (p+q)^2} \sum_{\lambda} 
\epsilon_{\nu}^{\ast (\lambda)}
\langle J/\psi(p-k, \epsilon^{(\lambda)}) K(q) | {\cal O}(0) | B(p+q) \rangle
 &=& \frac{1}{\pi^2\, m_{J/\psi} f_{J/\psi} }\nonumber \\ 
& & \hspace*{-12cm} \times \int_{4 m_c^2}^{s_0^{J/\psi}}
ds \frac{(m_{J/\psi}^2 + Q_0^2)^{n+1}}{(s + Q_0^2)^{n+1}} \int_{m_b^2}^{f_2(s,s_0^B,P^2,p^2)}  \frac{ds'}{s' - (p+q)^2}
{\rm Im}_{s'} {\rm Im}_s {\cal F}_{\nu}(s,s',P^2,p^2)                  
\label{eq:dd1}
\end{eqnarray}
and after the Borel transformation in $(p+q)^2$ variable, we can further write 
\begin{eqnarray}
\sum_{\lambda} \epsilon_{\nu}^{\ast (\lambda)}
\langle J/\psi(p-k, \epsilon^{(\lambda)}) K(q) | {\cal O}(0) | B(p+q) \rangle  &=& \frac{1}{\pi^2 \, m_{J/\psi} f_{J/\psi} m_{B}^2 f_{B}} \nonumber \\
& & \hspace*{-10cm}  \times \int_{4 m_c^2}^{s_0^{J/\psi}}
ds \frac{(m_{J/\psi}^2 + Q_0^2)^{n+1}}{(s + Q_0^2)^{n+1}} \int_{m_b^2}^{f_2(s,s_0^B,P^2,p^2)} ds' e^{(m_B^2 - s')/M^2}  
{\rm Im}_{s'} {\rm Im}_s {\cal F}_{\nu}(s,s',P^2,p^2) \, , 
\label{eq:dd2}
\end{eqnarray}                                                 
In above,  $M$ is the Borel parameter and the function $f_2$ is the upper limit of the $s$ integral after subtraction 
of continuum of $B$ channel. 

Further, to extract the kinematical structure of interests, 
we decompose  the matrix element $\langle J/\psi(p-k, \epsilon^{(\lambda)}) K(q) | 
 {\cal O}(0) | B(p+q) \rangle $ as
\begin{eqnarray}
\langle J/\psi(p-k, \epsilon^{(\lambda)}) K(q) | 
 {\cal O}(0) | B(p+q) \rangle 
&=& 
\epsilon \cdot q \, A^{(q)} + 
\epsilon \cdot k \,A^{(k)} \nonumber \\
& & \hspace*{-5cm}  +
\epsilon_{\alpha} \epsilon^{\alpha \rho \sigma \xi} 
(p-k)^{\rho} k^{\sigma} q^{\xi} \, A^{(\epsilon)}
\label{eq:ampD}
\end{eqnarray}
%In the above, it is assumed that the continuum in (\ref{eq:piB}) is already subtracted. 

By inserting this expansion in the expression (\ref{eq:dd2}), after 
the summation of the polarization vectors 
\begin{equation}
\sum_{\lambda} \epsilon_{\nu}^{\ast (\lambda)}  \epsilon_{\alpha}^{(\lambda)} =  
\left (-g_{\nu \alpha} + 
\frac{(p-k)_{\nu} (p-k)_{\alpha}}{(p-k)^2} \right )
\end{equation}
one obtains the sum rule for different kinematical structures:
\begin{eqnarray}
& & - k_{\nu} \;  A^{(k)} - q_{\nu} \; 
A^{(q)} - \epsilon_{\nu \rho \sigma \xi} (p-k)^{\rho} q^{\sigma} k^{\xi} \; A^{(\epsilon)} 
\nonumber \\
& &  \hspace*{0.5cm} 
+ (p - k)_{\nu} \left ( \frac{(p-k) \cdot k}{(p-k)^2} \; A^{(k)} + 
\frac{(p-k) \cdot q}{(p-k)^2} \;  A^{(q)} \right )  =  
\nonumber \\
& & \hspace*{0.3cm} = \frac{1}{\pi^2\,m_{J/\psi} f_{J/\psi}m_{B}^2 f_{B}}\int_{4 m_c^2}^{s_0^{J/\psi}} 
ds \frac{(m_{J/\psi}^2 + Q_0^2)^{n+1}}{(s + Q_0^2)^{n+1}} \int_{m_b^2}^{f_2(s,s_0^B,P^2,p^2)}  ds' e^{(m_B^2-s')/M^2}
\nonumber \\
& &  \hspace*{0.7cm} \times {\rm Im}_{s'} {\rm Im}_s \left [ k_{\nu} \;  F^{(k)} + q_{\nu} \;  
F^{(q)} + (p-k)_{\nu} F^{(p-k)} + 
\epsilon_{\nu \rho \sigma \xi} (p-k)^{\rho} q^{\sigma} k^{\xi} \;F^{(\epsilon)}
\right ] \nonumber \\
& &  \, . 
\label{eq:SRd}
\end{eqnarray}

The 
coefficient function in front of $(p-k)_{\nu}$ looks like 
\begin{eqnarray}
A^{(p-k)}  =   \left ( \frac{(p-k) \cdot k}{(p-k)^2} \; A^{(k)} +
 \frac{(p-k) \cdot q}{(p-k)^2} \;  A^{(q)} \right )  \, , 
\label{eq:Apk}
\end{eqnarray}
which is a consequence of the conserved  $J/\psi$ current. 
The sum rule expression for the $A^{(p-k)}$ part reads 
\begin{eqnarray}
A^{(p-k)} &=&  \frac{1}{\pi^2\,m_{J/\psi} f_{J/\psi}m_{B}^2 f_{B}}   
\nonumber \\
& & \hspace*{-2cm} \times \int_{4 m_c^2}^{s_0^{J/\psi}}
ds \frac{(m_{J/\psi}^2 + Q_0^2)^{n+1}}{(s + Q_0^2)^{n+1}} \int_{m_b^2}^{f_2(s,s_0^B,P^2,p^2)} 
ds' e^{(m_B^2 -s')/M^2} {\rm Im}_{s'} {\rm Im}_s \; F^{(p-k)}(s,s',P^2,p^2) \, . 
\nonumber \\
\label{eq:hadr3}
\end{eqnarray} 

At the end we analytically continue $P^2$ to $P^2 \ge 0$,  and 
choose $P^2 = m_B^2$. 
This enables the extraction of the physical matrix element, because 
the  unphysical momentum $k$ disappears from the 
ground state contribution, due 
to the simultaneous conditions applied, $P^2 = m_B^2$ and $(p+q)^2 = m_B^2$.  
From (\ref{eq:ampD}) and (\ref{eq:Apk}) follows
\begin{equation}
\langle J/\psi (p, \epsilon^{(\lambda)}) K(q) | {\cal O}(0) | B(p+q) \rangle = 
\epsilon \cdot q \, A^{(q)}(P^2 = m_B^2) = 
\epsilon \cdot q \frac{2 p^2}{m_B^2 - p^2} A^{(p-k)}(P^2 = m_B^2) 
\label{eq:amp1}
\end{equation}
and the final sum rule relation for 
the physical matrix element 
$\langle J/\psi (p, \epsilon^{(\lambda)}) K(q) | {\cal O}(0) | B(p+q) \rangle$ takes the 
form 
\begin{eqnarray}
\langle J/\psi (p, \epsilon^{(\lambda)}) K(q) | 
{\cal O}(0) | B(p+q) \rangle  &=& \nonumber \\
& & \hspace*{-7cm} =  2 \epsilon \cdot q m_{J/\psi} f_{J/\psi} 
\Bigg \{ \frac{1}{\pi^2  f_{J/\psi}^2} 
\int_{4 m_c^2}^{s_0^{J/\psi}} ds 
\frac{(m_{J/\psi}^2 + Q_0^2)^{n+1}}{(s + Q_0^2)^{n+1}} 
\frac{1}{m_{B}^2 f_{B}} \int_{m_b^2}^{f_2(s,s_0^B,m_B^2,p^2)} ds' e^{(m_B^2-s')/M^2}
\nonumber \\
& & \hspace*{-6.5cm} \times 
\left [ \frac{p^2}{\displaystyle m_{J/\psi}^2(m_B^2 - p^2)}\, 
{\rm Im}_{s'} {\rm Im}_s F^{(p-k)}
(s,s',m_B^2,p^2)  
\right ] \Bigg \} \, . 
\label{eq:amp}
\end{eqnarray}
%Here, the kinematical variable $p^2$ is still taken to be a free parameter. 

\end{widetext}

Some comments are in order. The $B \rightarrow J/\psi K$ case seems to be much more complicated than the 
decay of a $B$ meson to two light pions discussed in \cite{Khodja}. The complication does not 
appear only due to the massive $c$ quarks, or the vector structure of the $J/\psi$ current, but 
mainly due to the local duality assumption in $J/\psi$ channel, 
which is expected to work much worse than in the pion channel in the 
$B \rightarrow \pi \pi$ decay. Although it is 
possible to stay away from the the excited and resonant 
hadronic states in the $J/\psi$ channel, one can still expect that 
there will be an influence of the $\psi'$ resonance, which, in a more precise calculation has 
to be taken into account explicitly. 
The technical difficulties which are induced by the fact that the value of $P^2$ parameter is close 
to the 
hadronic threshold of $J/\psi$ are left for the discussion in Section~\ref{sec:calc}. 

\section{\label{sec:fact}Factorization in the light-cone sum rule approach}

We first consider the contribution of the ${\cal O}_2$ operator. As we have shown in the 
introduction, this operator 
contributes to the factorizable part of the matrix element $\langle J/\psi (p, \epsilon^{(\lambda)}) K(q) |
H_W | B(p+q) \rangle$. 

The main contribution comes from the diagram shown in Fig.2, where for ${\cal O} = {\cal O}_2$ 
there is no interaction 
between the charm loop and the $B-K$ system 
at the leading level. Therefore, the calculation of this contribution is rather simple. 
According to the expression (\ref{eq:amp}), the $(p-k)_{\nu}$ part of the 
correlation function (\ref{eq:corr0}), $F^{(p-k)}$ needs to be calculated and its double imaginary part 
has to be extracted.  
The calculation proceeds in several steps. One inserts first 
explicitly the $J/\psi$ and $B$ currents in (\ref{eq:corr0}), 
and takes the expression (\ref{eq:oper}) for the operator ${\cal O}_2$. 
The $c$-quarks are contracted to 
a $c \overline{c}$-loop and can be then independently integrated. 
The contraction of $b$-fields produces a free $b$-quark propagator and the rest of the fields 
is organized into the leading, 
twist-2 kaon distribution amplitude $\phi_K$. 
Explicitly, we obtain
\begin{equation}
F^{(p-k)}_{tw 2} = \frac{m_b^2 f_K}{4 \pi^2} \int_{4 m_c^2}^{\infty} \frac{ds}{s - (p-k)^2} q \cdot (p-k) \left 
( 1 + \frac{2 m_c^2}{s} \right ) \sqrt{1 - \frac{4 m_c^2}{s} } 
\int_0^1 du \frac{\phi_K (u)}{ m_b^2 - (p+q u)^2 } \, , 
\label{eq:facLCSR}
\end{equation}
where $\phi_K(u)$ is the kaon twist-2 distribution amplitude defined by 
\begin{equation}
\langle K(q)| \overline{s}(0) \gamma_{\mu} \gamma_5 u(x) | 0 \rangle = 
-i q_{\mu}\, f_K \int_0^1 du e^{i u q x} \phi_K(u) \, . 
\label{eq:facQCD}
\end{equation} 
%The other structure $F^{(k)}_{tw 2}$ does not show in the contribution the 
%${\cal O}_2$ operator. 

The first integral in (\ref{eq:facLCSR}), apart from the kinematical factor $q \cdot (p-k)$, is nothing else 
but the charm loop contribution to the vacuum polarization calculated in the sum 
rule approach \cite{Reinders}. 
The second integral, considered in the leading twist 
approximation, reduces exactly to the light-cone twist-2 expression 
for the $F_{B K}^+$ form factor \cite{BKR,KRWWY}. 
This part, with the substitution $ u = (m_b^2- p^2)/(s'-p^2)$, 
can be rewritten in a dispersion form as
\begin{equation}
\int_0^1 du \frac{\phi_K (u)}{ m_b^2 - (p+q u)^2 } = \int_{m_b^2}^{\infty} 
\frac{ds'}{s'-(p+q)^2} \frac{1}{s'-p^2} \phi_K(u(s')) \, . 
\end{equation}
In such a way the expression (\ref{eq:facLCSR}) receives the needed double dispersion form from 
which the double imaginary part in $s$ and $s'$ variables can be trivially extracted. 

The contribution of the ${\cal O}_2$ operator to the $B \rightarrow J\psi K$ matrix element then follows from the 
the sum rule relation (\ref{eq:amp}):
\begin{eqnarray}
\langle J/\psi (p, \epsilon^{(\lambda)}) K(q) |
{\cal O}_2(0) | B(p+q) \rangle_{tw 2} &=& \nonumber \\
& & \hspace*{-7cm} = 2 \epsilon \cdot q \frac{m_{J/\psi}}{f_{J/\psi}} 
\nonumber \\ & & \hspace*{-6.5cm}
\left [ 
\frac{(m_{J/\psi}^2+ Q_0^2)^{n+1}}{4 \pi^2}  
\int_{4 m_c^2}^{s_0^{J/\psi}} \frac{ds}{(s+ Q_0^2)^{n+1}} \left ( 1 + \frac{2 m_c^2}{s} \right ) 
\sqrt{1 - \frac{4 m_c^2}{s} } 
\frac{p^2}{m_{J/\psi}^2}\frac{\displaystyle (1-\frac{s}{m_B^2})}{\displaystyle (1-\frac{p^2}{m_B^2})} \right ] \nonumber \\ 
& & \hspace*{-5.5cm} \times \left [ \frac{f_K m_b^2}{2 f_B m_B^2} \int_{m_b^2}^{s_0^B} \frac{ds'}{s' - p^2}
e^{(m_B^2 - s')/M^2} \phi_K \left ( \frac{m_b^2-p^2}{s'-p^2} \right ) \right ]
\nonumber \\
& & \hspace*{-7cm}  \simeq 2 \epsilon \cdot q \, m_{J/\psi} f_{J/\psi} 
F_{B \rightarrow K}^+(p^2) \, . 
\label{eq:facfac}
\end{eqnarray}
Here we see that the amplitude $\langle J/\psi (p, \epsilon^{(\lambda)}) K(q) |
{\cal O}_2(0) | B(p+q) \rangle $ factorizes and
in a good approximation 
the factorizable expression for the \\
$\langle J/\psi (p, \epsilon^{(\lambda)}) K(q) |
{\cal O}_2(0) | B(p+q) \rangle $ amplitude, given by (\ref{eq:fac}), is recovered for 
$p^2 = m_{J/\psi}^2$.
%allowing numerically for the small variation around this value.                      
In the first parenthesis, apart from the small $(1-s/m_{B}^2)/(1-m_{J/\psi}^2/m_{B}^2)$ correction, 
there is the leading order expression for the $f_{J/\psi}^2$ in the QCD sum 
rule approach. The correction is the result of 
calculation with the nonvanishing $k$ momentum.  
The second parenthesis in (\ref{eq:facfac}) gives the twist-2 contribution to 
$F_{BK}^+(p^2)$ form factor. 
By reproducing the factorization result for $p^2 = m_{J/\psi}^2$, we fix the value 
$p^2 = m_{J/\psi}^2$ also in the further calculation.

\section{\label{sec:calc}Soft nonfactorizable contributions in the LCSR approach}

For a discussion of nonfactorizable contributions to the 
$B \rightarrow J/\psi K$ decay, we need to do a systematic 
$\alpha_s$ and twist expansion of the correlator (\ref{eq:corr0}). 

After explicit insertion of the interpolating $J/\psi$ and $B$ meson currents and the operator ${\cal O}_2$ or $\tilde{\cal O}_2$, 
the correlation function (\ref{eq:corr0}) can be written in form:
\begin{eqnarray}
F_{\nu}(p,q,k) &=& m_b \int d^4 x e^{-i(p+q)x} \int d^4 y e^{i(p-k)y} A^{jk} A^{lm} 
\nonumber \\
& & \hspace*{-1cm}\langle K(q)| Tr[\gamma_{\nu} S^{ij}(y,0|m_c)\Gamma_{\mu} S^{ki}(0,y|m_c)] 
\, \overline{s}^l \Gamma^{\mu} S^{mn}(0,x|m_b) \gamma_5 u^n | 0 \rangle \, , 
\end{eqnarray}
where $i,j,k,l,m,n$ are color indices, $S(x,y|m)$ are quark propagators defined in (\ref{eq:prop}) below 
and $A^{ij} = \delta^{ij}$ or $A^{ij} = T^{ij} = (\lambda^a/2)^{ij}$ 
for the insertion of ${\cal O}_2$ or $\tilde{\cal O}_2$ operator, respectively.  

The $\alpha_s$ and twist expansion is achieved by considering the light-cone expression for 
quark propagators.  
Up to terms proportional to $G$, the 
propagation of a massive quark in the external gluon field 
in the Fock-Schwinger gauge is given by \cite{BK}
\begin{eqnarray}
S^{ij}(x_1,x_2|m) &\equiv&  -i\langle 0 | T \{q^i(x_1)\, \bar{q}^j(x_2)\}| 0 \rangle
\nonumber
\\
%& & \equiv S^{(0)}(x_1,x_2,m) \delta^{ij} + g_s S^{(G)\, ij}(x_1,x_2,m) 
%\nonumber 
%\\
& & \hspace*{-1cm} = \int\frac{d^4k}{(2\pi)^4}e^{-ik(x_1-x_2)}\Bigg\{
\frac{\not\!k +m}{k^2-m^2} \delta^{ij}
-\int\limits_0^1 dv\,  g_s \, G^{\mu\nu}_a(vx_1+(1-v)x_2)
\left (\frac{\lambda^a}{2} \right )^{ij} 
\nonumber
\\
& & \hspace*{-0.5cm}\times \Big[ \frac12 \frac {\not\!k +m}{(k^2-m^2)^2}\sigma_{\mu\nu} -
\frac1{k^2-m^2}v(x_1-x_2)_\mu \gamma_\nu \Big]\Bigg\}\,. 
\label{eq:prop}
\end{eqnarray}
\begin{figure*}
\begin{center}
\includegraphics*[width=17cm]{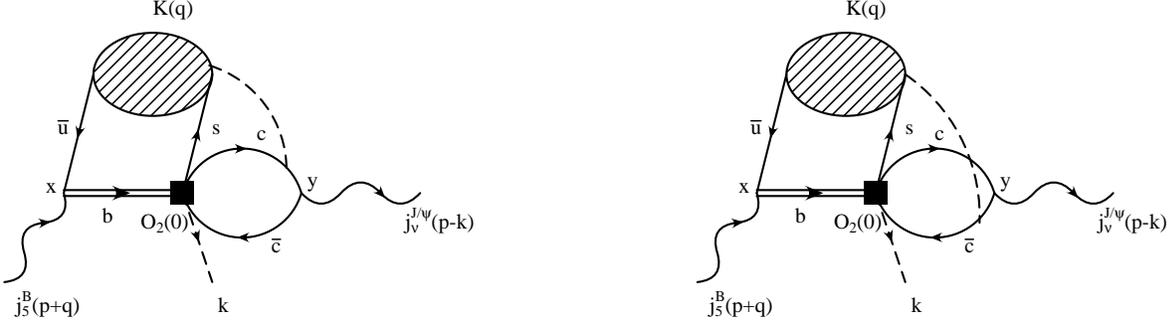}
\end{center}
\caption{Soft nonfactorizable contributions to the correlation function 
(\protect{\ref{eq:corr0}}).}
\end{figure*}
From the above, considering the color structure of the ${\cal O}_2$ operator, we can easily deduce that the nonfactorizable 
contribution from this operator appears first at the two-gluon level and is therefore of $O(\alpha_s^2)$.  
%Calculation of this contribution is technically very demanding and will not be discussed in this paper. 
On the contrary, nonfactorizable corrections from the $\tilde{\cal O}_2$ operator are already 
given by the one-gluon exchange. 
The leading hard nonfactorizable contributions are due to the exchange 
of a hard gluon between the $c$-quark (antiquark) and the one of the remaining $b$, $\overline{u}$ 
or $s$ quarks, see Fig.2. These 
contributions emerge at the two-loop level and although they are calculable in LCSR, their calculation 
is technically very demanding and will not be discussed in this paper. 

Insertion of the gluonic term of the propagator $S^{ij}(y,0,m_c)$ or $S^{ki}(0,y,m_c)$ yields the contributions represented 
in Fig.3. These are the leading soft nonfactorizable contributions. In terms of the light-cone expansion 
they are of the higher twist and described by the three  particle kaon distribution amplitudes defined 
by the following matrix elements: 
\\
- {\it  twist-3 distribution amplitude}
\begin{eqnarray}
\langle 0 |\overline{s}(0) \sigma_{\mu \nu} \gamma_5 G_{\alpha \beta}(v y) u(x) | K^{+}(q) \rangle &=& 
i f_{3 K} \left [ (q_{\alpha}q_{\mu}g_{\beta \nu} - q_{\beta}q_{\mu}g_{\alpha \nu}) \right .  \nonumber \\
& & \hspace*{-6cm} \left . - (q_{\alpha}q_{\nu}g_{\beta \mu} - q_{\beta}q_{\nu}g_{\alpha \mu}) \right ] 
\int {\cal D}\alpha_i \phi_{3 K}(\alpha_i, \mu) 
e^{-i q(x \alpha_1 + y v \alpha_3)}
\label{eq:tw3}
\end{eqnarray}
- {\it twist-4 distribution amplitudes}
\begin{eqnarray}
& & \langle 0 |\overline{s}(0) i\gamma_{\mu} \tilde{G}_{\alpha \beta}(v y) u(x) | K^{+}(q) \rangle = 
q_{\mu} \frac{q_{\alpha} x_{\beta} - q_{\beta}x_{\alpha}}{q x} f_{K} \int {\cal D}\alpha_i \tilde{\phi}_{||}(\alpha_i,\mu)
e^{-i q(x \alpha_1 + y v \alpha_3)} \nonumber \\
& & +
(g_{\mu \alpha}^{\perp}q_{\beta} - g_{\mu \beta}^{\perp}q_{\alpha}) \int {\cal D}\alpha_i \tilde{\phi}_{\perp}(\alpha_i,\mu)
e^{-i q(x \alpha_1 + y v \alpha_3)}
\label{eq:tw4a}
\end{eqnarray}
\begin{eqnarray}
& & \langle 0 |\overline{s}(0) \gamma_{\mu} \gamma_5 {G}_{\alpha \beta}(v y) u(x) |K^{+}(q) \rangle = 
q_{\mu} \frac{q_{\alpha} x_{\beta} - q_{\beta}x_{\alpha}}{q x} f_{K} \int {\cal D}\alpha_i {\phi}_{||}(\alpha_i,\mu)
e^{-i q(x \alpha_1 + y v \alpha_3)} \nonumber \\
& & +
(g_{\mu \alpha}^{\perp}q_{\beta} - g_{\mu \beta}^{\perp}q_{\alpha}) \int {\cal D}\alpha_i {\phi}_{\perp}(\alpha_i,\mu)
e^{-i q(x \alpha_1 + y v \alpha_3)} \, . 
\label{eq:tw4b}
\end{eqnarray}
In above, $\tilde{G}_{\alpha \beta} = \frac{1}{2} \epsilon_{\alpha \beta \rho \sigma} G^{\rho \sigma}$, 
$G^{\rho \sigma} = g_s \, \lambda^{a}/2 \, G^{\rho \sigma}_a$, ${\cal D} \alpha_i = 
d\alpha_1 d\alpha_2 d\alpha_3 \delta( 1 - \alpha_1 - \alpha_2 - \alpha_3)$, and $g_{\alpha \beta}^{\perp} = 
g_{\alpha \beta} - (x_{\alpha} q_{\beta} + x_{\beta} q_{\alpha})/qx$. 
Both twist-3 and twist-4 distribution amplitudes contribute at the same order. They are parameterized by  
\begin{eqnarray}
\phi_{3K}(\alpha_i,\mu) &=& 360 \alpha_1 \alpha_2 \alpha_3^2 \left (1 + a(\mu) \frac{1}{2} ( 7 \alpha_3 - 3) + b(\mu) 
(2 - 4 \alpha_1 \alpha_2 - 8 \alpha_3 (1 - \alpha_3)) \right . \nonumber \\
& &  + \left . c(\mu) ( 3 \alpha_1 \alpha_2 - 2 \alpha_3 + 3 \alpha_3^2) \right ) \, , 
\label{eq:amtw3} \\
\phi_{\perp}(\alpha_i,\mu) &=& 30 \delta^2(\mu)(\alpha_1-\alpha_2)\alpha_3^2\left [ \frac{1}{3} + 
2 \epsilon (\mu) (1 - 2 \alpha_3) \right ]  \, , 
\label{eq:amtw4a} \\
\phi_{||}(\alpha_i,\mu) &=& 120 \delta^2(\mu) \epsilon (\mu)  (\alpha_1-\alpha_2) \alpha_1 \alpha_2 \alpha_3  \, , 
\label{eq:amtw4b} \\
\tilde{\phi}_{\perp}(\alpha_i,\mu) &=& 
30 \delta^2(\mu) \alpha_3^2 ( 1 - \alpha_3) \left [ \frac{1}{3} + 2 \epsilon (\mu)  (1 - 2 \alpha_3) \right ] \, , 
\label{eq:amtw4ta} \\
\tilde{\phi}_{||}(\alpha_i,\mu) &=& -120 \delta^2(\mu) \alpha_1 \alpha_2 \alpha_3 \left [ \frac{1}{3} + 
\epsilon (\mu) (1 - 3 \alpha_3) \right ] \, . 
\label{eq:amtw4tb} \
\end{eqnarray}
The parameters are estimated from sum rules \cite{CZ,BF} and the values are listed in \cite{KR2}. 
In the numerical evaluation we use the asymptotic form of the above expressions where 
$a(\mu),b(\mu), c(\mu)$ and $\epsilon(\mu)$ dependence is neglected. The asymptotic expressions 
for twist-3 and twist-4 distribution amplitudes should provide sufficiently reliable estimates of 
already subleading contributions. 

The QCD calculation of two diagrams in Fig.3 at the twist 3 level 
yields 
\begin{eqnarray}
F^{(p-k)}_{tw 3} &=& - \frac{m_b f_{3 K}}{4 \pi^2} \int_0^1 dv \int {\cal D}\alpha_i \frac{\phi_{3K}(\alpha_i, \mu)}
{m_b^2 - (p + q (1-\alpha_1))^2} \, 
\int_0^1 dx \frac{ 2 x^2 (1-x)}{ m_c^2 - \overline{Q}^2 x (1-x)  
} \nonumber \\
& & \times \, q \cdot (p-k) \Big [ (2-v) q \cdot k + 2 (1-v) q \cdot (p-k) \Big ] \, , 
\label{eq:tw3c}
\end{eqnarray}
where $\overline{Q} = p-k+v \alpha_3 q$. 
Comparing the above expression with the one obtained for the $B \rightarrow \pi \pi$ case, 
(Eq. (26) in \cite{Khodja}), 
we can see that there is an additional, $x$-integral for the massive $c \overline{c}$ loop. 
Otherwise, the expressions are the same and for $m_c \rightarrow 0$ the result form  Eq. (26) in \cite{Khodja} is 
exactly recovered, up to a sign, which can be traced back to a difference between the 
pseudoscalar and vector currents interpolating $\pi$ and $J/\psi$, respectively.
%is a consequence of dealing with the vector current here, 
%and which is compensated by the different signs in the sum rule expressions for these 
%two cases. 

By changing the order and variables of integration 
one can bring
the above expression into the following form:
\begin{eqnarray}
F^{(p-k)}_{tw 3} &=&  \frac{m_b f_{3 K}}{16 \pi^2} \int_{4 m_c^2}^{\infty} \frac{ds}{s - (p-k)^2}
\int_0^{1-\frac{4 m_c^2}{s}} \frac{dy}{2 \sqrt{y}} \int_{x(s,y,P^2)}^1 \frac{du}{m_b^2 - (p+q u)^2} 
\nonumber \\
& & \times \int_{x(s,y,P^2)}^u 
\frac{dv}{v^2} \phi_{3K}(1 -u, u-v, v) 
\nonumber \\
& & \hspace*{1cm} \times 
\Big [ s - \frac{4 m_c^2}{1-y} +   
\left ( (p+q)^2 - p^2 \right ) 
( 2 v - x(s,y,P^2) ) \Big  ] \, , 
\label{eq:tw3pk}
\end{eqnarray}
and 
\begin{equation}
x(s,y,P^2) = \frac{s - \frac{\displaystyle 4 m_c^2}{\displaystyle 1-y^2}}{s-P^2}  \, .
\end{equation}

It is important to emphasize here that the above expression (\ref{eq:tw3pk}) 
is defined only for large spacelike momentum $|P^2| \sim m_b^2$.  
Furthermore, the expression (\ref{eq:tw3pk}) does not have a needed double dispersion form. 

In order to proceed we write 
\begin{eqnarray}
F^{(p-k)}_{tw 3} &=& \frac{1}{\pi} \int_{4 m_c^2}^{\infty} \frac{ds}{s - (p-k)^2} 
{\rm Im_s}\, F_{tw3}^{(p-k)}
(s, (p+q)^2, P^2,p^2) \, , 
\label{eq:tw3im1}
\end{eqnarray}
where 
\begin{eqnarray}
{\rm Im_s}\, F_{tw3}^{(p-k)}
(s, (p+q)^2, P^2,p^2)
&=& \nonumber \\
& & \hspace*{-5.5cm}   \frac{m_b f_{3 K}}{16 \pi} 
\int_0^{1-\frac{4 m_c^2}{s}} \frac{dy}{2 \sqrt{y}} \int_{x(s,y,P^2)}^1 \frac{du}{m_b^2 - (p+q u)^2} 
\int_{x(s,y,P^2)}^u 
\frac{dv}{v^2} \phi_{3K}(1 -u, u-v, v) 
\nonumber \\
& & \hspace*{-5cm} \times \Big [ s - \frac{4 m_c^2}{1-y} + 
\left ( (p+q)^2 - p^2 \right ) 
( 2 v - x(s,y,P^2) ) \Big  ] \, . 
\label{eq:tw3im2}
\end{eqnarray}

Now, it is possible to 
use the quark-hadron duality in $J/\psi$ channel and to 
subtract the $J/\psi$ continuum states by approximating them by (\ref{eq:tw3im2}), 
which changes the upper limit of $s$ integration in (\ref{eq:tw3im1}) to 
$s_0^{J/\psi} \sim 15 \, {\rm GeV^2}$.  This restriction of the $s$ integration enables the expansion of the imaginary part 
${\rm Im}_s F_{tw3}$ in the $x(s,y,P^2)$ variable. To reach the 
satisfactory precision we expand (\ref{eq:tw3im2}) up to order $O(x^3)$:
\begin{eqnarray}
{\rm Im_s} F_{tw3}^{(p-k)}(s,(p+q)^2,P^2) &=& 
 \nonumber \\
 & & 
 \hspace*{-5cm} \frac{m_b f_{3 K}}{16 \pi^2} \int_0^1 \frac{du}{m_b^2 - (p+q u)^2} 
 \int_0^{1-\frac{4 m_c^2}{s}} \frac{dy}{2 \sqrt{y}} 
\nonumber \\
& & \hspace*{-5cm} \Bigg\{
 \int_0^u
 \frac{dv}{v^2} \phi_{3K}(1 -u, u-v, v) 
 \left [ s -  \frac{4 m_c^2}{1-y} + 2 v ((p+q)^2 - p^2) \right ] \nonumber \\
 & & \hspace*{-4cm}  
 - \left [ \int_0^u \frac{dv}{v^2} \phi_{3K}(1 -u, u-v, v) \left ( (p+q)^2 - p^2 \right ) 
\right . \nonumber \\
 & & \hspace*{-4cm}  \left .  + \left ( s -  \frac{4 m_c^2}{1-y} \right ) \left ( \frac{1}{v^2} \phi_{3K}(1 -u, u-v, v) \right )_{v=0} 
 \right ] x(s,y,P^2) \nonumber \\
 & &  \hspace*{-5cm} - 
  \left ( s -  \frac{4 m_c^2}{1-y} \right ) 
 \left [ \frac{\partial}{\partial v} \left ( \frac{1}{v^2} \phi_{3K}(1 -u, u-v, v) \right ) \right ]_{v=0}  
 \frac{x^2(s,y,P^2)}{2} \Bigg\}
  +  O(x^3) \, . 
 \label{eq:tw3exp}
 \end{eqnarray}
%In the last integral we have already taken the explicit form of $ \phi_{3K}(1 -u, u-v, v)$ function 
%and have performed the $v$ integration for the simplicity.

In the above expression it is important to keep in mind that $s$ receives values 
in the range $4 m_c^2 < s < s_0^{J/\psi}$. 
It has also to be noted that the coefficients  in the expansion are $P^2$ independent objects. 
%$Im_s F_{tw3}^{(i)}$, $i=0,1,2,..$ are $P^2$ independent object, 
So, although, the above expression was derived for $P^2 < 0$,  the 
complete expression (\ref{eq:amp}) for the physical amplitude $B \rightarrow J/\psi K$  
is an analytic function in $P^2$ and it can be analytically continued to the positive values of $P^2 = m_B^2$. 
The result is more reliable for smaller $O(s/P^2)$ corrections. 
%that the corrections are relatively small. 
In our case, although the expansion is well converging, the first order correction in $\sim x(s,y,P^2)$ 
amounts to 
$ \sim 25\%$, which is 
significantly larger than the similar correction $\sim s_0^{\pi}/P^2 \sim O(1\, {\rm GeV}/m_B^2)$
in the 
$B \rightarrow \pi \pi $ case. 
Therefore, in the calculation of the soft nonfactorizable
correction for $B \rightarrow J/\psi K$, the analytical continuation of $P^2$ to its positive value 
embeds an unavoidable theoretical uncertainty. 
However, $O(x^2)$ corrections are already at a percent level, and 
the expansion is well converging. 

The same procedure employed for twist-4 contributions gives somewhat more 
complicated result:
\begin{eqnarray}
F^{(p-k)}_{tw 4} &=& \frac{m_b^2 f_{K}}{8 \pi^2} \int_{4 m_c^2}^{\infty} \frac{ds}{s - (p-k)^2}
\int_0^{1-\frac{4 m_c^2}{s}} \frac{dy}{2 \sqrt{y}} \int_{x(s,y,P^2)}^1 
\frac{du}{m_b^2 - (p+q u)^2} 
\nonumber \\
& & \hspace*{-1cm}\times \int_{x(s,y,P^2)}^u 
\frac{dv}{v} \tilde{\phi}_{\perp}(1 -u, u-v, v) 
\left [ 3 - \frac{2}{v} x(s,y,P^2) \right ] 
\nonumber \\
& &  \hspace*{-2cm} + \frac{m_b^2 f_{K}}{8 \pi^2} \int_{4 m_c^2}^{\infty} \frac{ds}{s - (p-k)^2}
\int_0^{1-\frac{4 m_c^2}{s}} \frac{dy}{2 \sqrt{y}} \int_{x(s,y,P^2)}^1 
\frac{du}{\left [ (m_b^2 - (p+q u)^2 \right ]^2}
\nonumber \\
& & \hspace*{-1cm}\times \int_{x(s,y,P^2)}^u
\frac{dv}{v^2} \Phi_{1}(1 -u, v)
\left [ s - \frac{4 m_c^2}{1-y} + ((p+q)^2 - p^2) ( -v + x(s,y,P^2) )\right ] 
\nonumber \\
& & \hspace*{-2cm} - \frac{m_b^2 f_{K}}{8 \pi^2} \int_{4 m_c^2}^{\infty} \frac{ds}{s - (p-k)^2}
\int_0^{1-\frac{4 m_c^2}{s}} \frac{dy}{2 \sqrt{y}} \int_{x(s,y,P^2)}^1 
\frac{du}{\left [m_b^2 - (p+q u)^2 \right ]^2}
\nonumber \\
& & \hspace*{-1cm}\times 
\frac{\Phi_{2}(u)}{u^2}
\left [ s - \frac{4 m_c^2}{1-y} + ((p+q)^2 - p^2) (-u + x(s,y,P^2) \right ]
\nonumber \\
& & \hspace*{-2cm} - \frac{m_b^2 f_{K}}{8 \pi^2} \int_{4 m_c^2}^{\infty} 
\frac{ds}{\left [ s - (p-k)^2 \right ]^2}
\int_0^{1-\frac{4 m_c^2}{s}} \frac{dy}{2 \sqrt{y}} \frac{1-y}{1 -y - \frac{4m_c^2}{P^2}}
%\nonumber \\
%& & \hspace*{-1cm} \times 
\int_{x(s,y,P^2)}^1 
\frac{du}{m_b^2 - (p-q u)^2}
\nonumber \\
& & \hspace*{-1cm} \times 
\frac{\Phi_{2}(u)}{u^2}
\left [ \frac{x(s,y,P^2)}{-P^2}\left (2\, q \cdot (p-k) \right )^2 
\left (1 -\frac{x(s,y,P^2)}{u}\frac{q\cdot k}{q \cdot p}\right ) \right ] \, . 
\label{eq:tw4pk}
\end{eqnarray}
Here, $ q\cdot k = \frac{1}{2}( (p-k)^2 - P^2 + (p+q)^2 - p^2 )$ and 
$q \cdot p = \frac{1}{2}( (p+q)^2 -p^2))$. 

The twist-4 wave functions appear in combinations 
\begin{eqnarray}
\Phi_1(u,v) &=& \int_0^u d\omega \left ( 
\tilde{\phi}_{\perp}(\omega, 1 - \omega - v,v) + \tilde{\phi}_{||}(\omega, 1 - \omega - v,v) \right )
\nonumber \\
\hspace*{-0.5cm} \Phi_2(u) &=& \int_0^u d\omega' \int_0^{1-\omega'} d \omega'' \left ( 
\tilde{\phi}_{\perp}(\omega'', 1 - \omega'' -\omega', \omega') + 
\tilde{\phi}_{||}(\omega'', 1 - \omega'' -\omega', \omega') \right )
\label{eq:tw4wf}
\end{eqnarray}
 
The first term in (\ref{eq:tw4pk}) can be treated in a similar 
way as the twist-3 part, $F_{tw3}^{(p-k)}$, expanding in $x(s,y,P^2)$ with the result 
\begin{eqnarray}
F_{tw4}^{(p-k)} &=&
\frac{m_b^2 f_K}{8 \pi^2} \int_{4 m_c^2}^{s_0^{J/\psi}} \frac{ds}{s-(p-k)^2} 
\int_0^1 \frac{du}{m_b^2 - (p+q u)^2}
\int_0^{1-\frac{4 m_c^2}{s}} \frac{dy}{2 \sqrt{y}} 
\nonumber \\
& & \hspace*{-1cm} \times 
\int_0^u
\frac{dv}{v^2} \tilde{\phi}_{\perp}(1 -u, u-v, v)
\left [ 3 -   \frac{2}{v} x(s,y,P^2) \right ]  
 +  O(x^3) \, .
\label{eq:tw4exp}
\end{eqnarray}

Other parts in (\ref{eq:tw4pk}) contain denominators 
of a form
\begin{equation}
\frac{1}{\left [ s - (p-k)^2 \right ]^2} \qquad {\rm or}\qquad  
\frac{1}{\left [ m_b^2 - (p+u q)^2 \right ]^2} \, , 
\end{equation}
which are typical for twist-4 contributions.

To be able to deal with such terms we perform a  partial integration.
However, the problem is the subtraction of a continuum for such terms, 
because the complete expression does not possess the needed dispersion form, where the hadronic 
spectral density can be 
identified with the imaginary part of the QCD amplitude, unless the surface terms  
are equal to zero.  
Fortunately, twist-4 contributions with the 
higher power of denominators numerically appear to be suppressed. Their contribution, 
neglecting the surface terms, is in the region of a few percent. 
Uncertainties involved in the LCSR calculation are certainly much 
larger, and we argue that the contributions with the higher power of denominators in 
the twist-4 part can be safely neglected in the numerical calculation. 

It is important to emphasize that 
due to the specific configuration of momenta, imposed by the 
$J/\psi$ continuum subtraction, the analytical continuation of
$P^2$ does not produce an imaginary phase. From this continuation, 
one would expect to get an imaginary phase in the penguin contributions of operators 
${\cal O}_1$ and ${\cal O}_2$. The phase 
is typical for such kind of contributions and known as the BSS phase \cite{BSS}. 
%because in 
%diagrams with penguin insertions, $P^2$ is not restricted to
%$|P^2| < s_0^{J/\psi}$. 
However, the penguin contributions in the process 
under the consideration are suppressed in 
the large $N_c$ limit by $1/N_c$ (additionally to the $1/N_c$ suppression of the 
emission amplitude calculated here), and are beyond a scope of this calculation. 

Putting twist-3, eqs.(\ref{eq:tw3im1}) and (\ref{eq:tw3exp}), and twist-4 (\ref{eq:tw4exp}) expressions together and subtracting the continuum of 
$B$ states, 
the final expression for the soft contributions to the $B \rightarrow J/\psi K$ amplitude,  
in the approximations discussed above, has the form 
\begin{eqnarray}
\langle J/\psi (p, \epsilon^{(\lambda)}) K(q) |
{\cal O}(0) | B(p+q) \rangle  &=& 
 2 \epsilon \cdot q \, m_{J/\psi} f_{J/\psi} \cdot
\nonumber \\
& & \hspace*{-7cm}
\frac{1}{ 4 \pi^2  f_{J/\psi}^2}
\int_{4 m_c^2}^{s_0^{J/\psi}} ds
\frac{(m_{J/\psi}^2 + Q_0^2)^{n+1}}{(s + Q_0^2)^{n+1}}
\frac{1}{2 m_{B}^2 f_{B}} \int_{u_0^B}^1 \frac{du}{u} e^{(m_B^2-(m_b^2 - m_{J/\psi}^2(1-u))/u)/M^2}
\nonumber \\
& & \hspace*{-6.5cm} \times
\int_0^{1-\frac{4 m_c^2}{s}} \frac{dy}{2 \sqrt{y}} \, \frac{m_b}{m_B^2 -m_{J/\psi}^2} 
\Bigg \{ \nonumber \\
& & \hspace*{-6cm} \frac{f_{3K}}{2} \Bigg [ 
 \int_0^u
 \frac{dv}{v^2} \phi_{3K}(1 -u, u-v, v)
 \left ( \frac{m_b^2 - m_{J/\psi}^2}{u} ( 2 v -  x(s,y,m_B^2) ) + s -  \frac{4 m_c^2}{1-y} \right )  \nonumber \\
 & & \hspace*{-5cm}  - \left ( s -  \frac{4 m_c^2}{1-y} \right ) \left ( \frac{1}{v^2} \phi_{3K}(1 -u, u-v, v) \right )_{v=0}
  x(s,y,m_B^2) \nonumber \\
 & &  \hspace*{-5cm} -
  \left ( s -  \frac{4 m_c^2}{1-y} \right )
 \left [ \frac{\partial}{\partial v} \left ( \frac{1}{v^2} \phi_{3K}(1 -u, u-v, v) \right ) \right ]_{v=0}
 \frac{x^2(s,y,m_B^2)}{2} \Bigg ]
\nonumber \\
& & \hspace*{-5cm} +
m_b f_K \int_0^u
\frac{dv}{v^2} \tilde{\phi}_{\perp}(1 -u, u-v, v)
\left [ 3 -   \frac{2}{v} x(s,y,m_B^2) \right ] 
 \Bigg \} \, ,
\label{eq:finalresult}
\end{eqnarray}
where $u_0^B = (m_b^2 - m_{J/\psi}^2)/(s_0^B - m_{J/\psi}^2)$.

\section{\label{sec:num}Numerical predictions}

Before giving numerical predictions on the soft nonfactorizable contributions, we have first to 
specify the numerical values of the parameters used. 

For parameters in the $B$ channel we use $m_B =  5.1$ GeV and the values taken from \cite{BKR}:
$f_B = 180 \pm 30 $ GeV, $m_b = 4.7 \pm 0.1$ GeV, and $s^{B}_0 = 35 \pm 2\, {\rm GeV}^2$. 
For $J/\psi$ we use the following: 
$m_{J/\psi} = 3.5$ GeV, $f_{J/\psi} = 0.405  \pm 0.014$ GeV from Eq.(\ref{eq:fpsi}), $m_c = 1.25 \pm 0.1$ and 
$s^{J/\psi}_0 = 15 \pm 2\, {\rm GeV}^2$ \cite{Reinders}. The $K$ meson decay constant is taken 
as $f_K = 0.16$ GeV.  For parameters which enter the coefficients of the twist-3 and 
twist-4 kaon wave functions we suppose that $f_{3 \pi} \simeq f_{3 K}$ and 
$\delta^2_K \simeq \delta_{\pi}^2$,  
and take $f_{3 K} = 0.0026$ GeV, $\delta^2(\mu_b) = 0.17$ GeV, where 
$\mu_b = \sqrt{m_B^2 - m_b^2} \sim m_b/2 \sim 2.4$ GeV \cite{CZ,BF}.  
 
Like in any sum rule calculation it is important that the stability criteria 
for (\ref{eq:amp}) are established by finding the window in $n$ and $M^2$ parameters in which, 
on the one hand, excited and continuum states are suppressed and on the 
other hand, a reliable 
perturbative QCD calculation is possible. 
The stability region for the Borel parameter is found in the interval $M^2 = 10 \pm 2\, {\rm GeV}^2$,  
known also from the other LCSR calculation of $B$ meson properties. 
Concerning moments in $J/\psi$ channel, the calculation is rather stable on the change of $n$ in 
the interval $n = 4 - 6$. 
$Q_0^2$ is parameterized by $Q_0^2 = 4 m_c^2 \xi$,
where $\xi$ is usually allowed to take values from 0 to 1.
As it was argued in \cite{Reinders}, where sum rules was applied for calculating the mass of $J/\psi$,  
and was also observed in our calculation, at $Q_0^2=0$ ($\xi = 0$) there is essentially no
stability plateau where $n$ is small enough that the QCD result is reliable and at the same
time the lowest lying resonance dominates. More stable result is achieved for $\xi \neq 0$. 
However, the result appears to be sensitive at most to the variation of the parameters 
$s^{B}_0$ and $s^{J/\psi}_0$. 

The numerical results for the soft nonfactorizable contributions are as follows
\begin{eqnarray}
\tilde{F}_{BK, \, tw3}^+(\mu_b) = 0.0051  \, , \qquad
\tilde{F}_{BK, \, tw4}^+(\mu_b) = 0.0089 \, , 
\end{eqnarray}
calculated at the typical $\mu_b \sim m_b/2$ scale of LCSR calculation. 
The above values are obtained for $n = 5$, $M_B^2 = 10\, {\rm GeV}^2$ and
$\xi = 0.5$. 
In general, one could expect that twist-4 contributions are relatively $O(1/m_b)$ suppressed
with respect to the twist-3 part and therefore are smaller.
However, careful study of the heavy-quark
mass behavior of the final expression (\ref{eq:finalresult}) shows that 
in the heavy-quark limit the twist-3 and twist-4 contributions are of the same
order (and are both suppressed by $1/m_b$ with respect to the factorizable
part (\ref{eq:facfac})). Therefore, it is not surprising that the numerical
contribution of the twist-4 part is relatively large.
Even- and odd-twist contributions stem 
from different chiral structures of the $b$-quark propagator and are, therefore, independent.
The $1/m_b$ suppression should, however, certainly be true when we compare
even(odd)-twist contributions among themselves (i.e. the twist-4 with the twist-2 contribution; 
the twist-5 with the twist-3 part etc. ). 

The variation of the sum
rule parameters implies the values:
\begin{eqnarray}
\tilde{F}_{BK, \, tw3}^+(\mu_b) = 0.004-0.007 \, , \qquad
\tilde{F}_{BK, \, tw4}^+(\mu_b) = 0.006-0.012
\end{eqnarray}
and the final value
\begin{eqnarray}
\tilde{F}_{BK}^+(\mu_b) = 0.011-0.018 \, . 
\label{eq:fth}
\end{eqnarray}

First, we note that the nonfactorizable part (\ref{eq:fth}) is much 
smaller than the $B-K$ transition form factor (\ref{eq:f0}) which enters the factorization result. 
It is also significantly 
smaller than its value (\ref{eq:fexp2}) extracted from experiments. Nevertheless, its influence on the 
final prediction for $a_2$ is significant, because of the large coefficient $2 C_1$ multiplying 
it. Furthermore, one has to emphasize that $\tilde{F}_{BK}^+$ is a positive 
quantity. Therefore, we do not find a theoretical support for the large $N_c$ limit 
assumption discussed in Section~\ref{sec:intro}, that the factorizable part proportional to $C_1(\mu)/3$ 
should at least be partially canceled by the nonfactorizable part. Our result also 
contradicts the result of the earlier application of QCD sum rules to 
$B \rightarrow J/\psi K$ \cite{KR2}, where negative and somewhat larger value for 
$\tilde{F}_{BK}^+$ was found. However, earlier applications  
of QCD sum rules to 
exclusive $B$ decays exhibit some deficiencies discussed in \cite{Khodja}.
In \cite{KR2}, mainly the problem was the separation of the ground state contribution in the B-channel
and the wrong $m_b \to \infty$ limit of higher-twist
terms obtained by using the short-distance expansion of the four-point
correlation function. In this work, following
the procedure taken from \cite{Khodja},
the problem is solved by introducing
the auxiliarly momentum $k$ in the $b$-decay vertex and by applying
the QCD light-cone sum rules.

Using the same values for the NLO Wilson coefficients as in Section~\ref{sec:fac}, one gets from (\ref{eq:fth}) 
for the effective coefficient $a_2$ the following value
\begin{equation}
a_2 \sim 0.15 -0.18\, |_{\mu = \mu_b} \, .
\label{eq:a2calc}
\end{equation}

Although the soft correction contribute at the order of $\sim 30 \% - 70\%$, the net result 
(\ref{eq:a2calc}) is still by 
approximately factor of two smaller than the experimentally determined value (\ref{eq:a2exp}). 

\section{\label{sec:comp}QCD factorization for the $B \rightarrow J/\psi K$ decays and
the impact of soft nonfactorizable corrections}

In an expansion in $1/m_b$ and $\alpha_s$, matrix elements for some of two-body decays of a $B$ meson 
can be computed consistently by the QCD factorization method \cite{BBNS}. 
This model applied to 
the $ B \rightarrow J/\psi K$ decay gives
\begin{eqnarray}
\langle J/\psi K | {\cal O} | B \rangle &=&
\langle K| \overline{s}\Gamma_{\mu} b| B \rangle \langle J/\psi|\overline{c}\Gamma^{\nu} c|0 \rangle
\left [ 1 + O(\alpha_s) +
O \left (\frac{\Lambda_{QCD}}{m_b} \right ) \right ]
\nonumber \\
&=& F_{BK}(m_{J/\psi}^2) \int_0^1 T^I(u) \phi_{J/\psi}(u)
\nonumber \\ & & + \int d \xi du dv T^{II}(\xi,u,v) \phi_B(\xi) \phi_{K}(v) 
\phi_{J/\psi}(u)  + O \left (\frac{\Lambda_{QCD}}{m_b} \right ) \, . 
\label{eq:factJK}
\end{eqnarray}
$T^I$ and $T^{II}$ are
perturbatively calculable hard scattering kernels and $\phi_{M=B,K,J/\pi}$ are meson light-cone 
distribution amplitudes. $T^I$ starts at order $O(\alpha_s^0)$, and at higher order of $\alpha_s$ contains 
nonfactorizable corrections from hard gluon exchange or penguin topologies. 
Hard nonfactorizable corrections in which the spectator of $B$ meson contribute are isolated in $T^{II}$.  
Soft nonfactorizable corrections denoted above as $ O(\Lambda_{QCD}/m_b)$  effects cannot be
calculated in the QCD factorization approach. According to some general considerations 
\cite{BBNS} these effects are expected to be suppressed, but there is no real confirmation of this conclusion.

In the limit $m_c \sim m_b \rightarrow \infty$, it can be shown \cite{BBNS} that at 
the leading order in $1/m_b$
there is no long distance interactions between $J/\psi$ and the rest $B- K$ system 
and the factorization holds. 
Actually, the $J/\psi$ case is
somewhat exceptional, since  
soft gluons in this limit are suppressed only by a factor $\Lambda_{QCD}/(m_b \alpha_s)$ \cite{BBNS} rather
than by $\Lambda_{QCD}/m_b$ like, for example, in the $B \rightarrow D \pi$ decay, for which the factorization has be 
proved at the two-loop level. 
If $J/\psi$ is treated as a light meson relative to $B$, then the 
factorization is recovered at $m_c/m_b \rightarrow 0$ limit. 
Unfortunately, for the
higher $1/m_b$ corrections, the factorization breaks down \cite{BBNS}.  

In connection to (\ref{eq:factJK}) the following should be emphasized. 
In the heavy quark limit $m_b \rightarrow \infty$ the hard scattering kernel $T^I$ is nothing else but the
$J/\psi$ meson decay constant and by neglecting $\alpha_s$ and $O(\Lambda_{QCD}/m_b)$ 
corrections, the naive factorization
result (\ref{eq:fac}) is recovered.  
%The transition form factor $F_{BK}(m_{J/\psi}^2)$ 
%%represents the nonperturbative input in the QCD factorization approach and
%its value is usually taken from the light-cone sum rule calculation. 
%iii) the hard scattering kernel $T^{II}$ which arise from the interaction of $J/\psi$ with
%the spectator quark of the $B$ meson should be taken into account 
%%cannot be neglected, especially for the higher twist kaon distribution 
%amplitudes $\phi_K$ \cite{Cheng};
In the hard corrections appear 
$J/\psi$ and $B$ meson
light-cone distribution amplitudes. Under the assumption that $m_c \ll m_b$, the 
%leading twist 
light-cone distribution amplitudes for $J/\psi$ can be taken to be equal to that of the 
$\rho$ meson, as it was done in \cite{Cheng} 
(vector meson distribution amplitudes were elaborated in \cite{BB}), although this assumption is not 
completely justified. 
However, we cannot say much about the 
$B$ meson distribution amplitude, except that it can be modeled or extracted form the 
experimental data \cite{LiM}, which is again model dependent. 
Fortunately, after some simplification, the result depend only on the
first moment of the $\phi_B$, $\int_0^1 d\xi \phi_B(\xi)/\xi = m_B/\lambda_B$, and therefore there is
a need for fixing just one parameter, 
$\lambda_B$. There is not much known about this parameter, except
its upper bound, $3 \lambda_B \le 4 (m_B - m_b)$, or effectively, $\lambda_B < 600$ MeV \cite{KPY}.

Here, we would like to discuss our results for the soft nonfactorizable contributions in 
comparison with the hard
nonfactorizable effects calculated in QCD factorization approach. 
As it was already noted in \cite{Khodja}, in the heavy quark limit the soft nonfactorizable 
contributions are suppressed by $1/m_b$ in comparison to the twist-2 factorizable part, which 
confirms the expansion in (\ref{eq:factJK}). 
With the inclusion of the hard nonfactorizable corrections, the $a_2$ parameter (\ref{eq:a2def}) appears 
as follows
\begin{equation}
a_2 = C_2(\mu) + \frac{C_1(\mu)}{3} +
2 C_1(\mu)
\left [ \alpha_s F^{hard}(\mu) +  
\frac{\tilde{F}_{BK}^+(\mu)}{F_{BK}^+(m_{J/\psi}^2)} \right ]   \, .
\label{eq:a2defN}
\end{equation}

The hard nonfactorizable contribution $F^{hard}$ were calculated in \cite{Cheng}.
The analysis was done up to twist-3 terms for the $K$ meson wave function which enters the
calculation of the $T^{II}$ hard scattering kernel in (\ref{eq:factJK}).
It is a well known feature of QCD factorization
that it breaks down by inclusion of higher-twist effects. The hard scattering kernel $T^{II}$ becomes logarithmically
divergent, which signalizes that it is dominated by the soft gluon exchange between the constituents of the 
$J/\psi$ and the spectator quark in the $B$ meson. In the QCD factorization
this logarithmic divergence  is usually parameterized by some
arbitrary complex parameter $r$ as $\int_0^1 dv/v = \ln(m_B/\Lambda_{QCD}) + r$ and although it is 
suppressed by $1/m_b$, this contribution is chirally enhanced
by a factor $2 m_K^2/((m_s +m_u))$.  This large correction makes it dangerous to take the
estimation for the twist-3 contribution literally, due to the possible large uncertainties which 
the parameter $r$ bears with.

The estimation done in the QCD factorization \cite{Cheng} shows hard-gluon 
exchange corrections to the naive factorization result of the order of $\sim 25 \%$, 
predicted by the LO calculation with the twist-2 kaon distribution 
amplitude. Unlikely large corrections are obtained by inclusion 
of the twist-3 kaon distribution amplitude. Anyhow, 
due to the obvious dominance of soft contributions to the twist-3 
part of the hard corrections in the QCD factorization \cite{BBNS}, 
it is very likely that some double counting of soft 
effects could appear if we naively compare the results.  
Therefore, taking only the twist-2 hard nonfactorizable corrections from \cite{Cheng} into 
account, recalculated at the $\mu_b$ scale, our  prediction (\ref{eq:a2calc}) changes to
\begin{equation}
|a_2| = 0.17 - 0.19 \, |_{\mu = \mu_b}
\label{eq:a2final}
\end{equation}
The prediction still remains to be too small to explain the data. 

Nevertheless, there are several things which have to be 
stressed here. Soft nonfactorizable contributions are at least equally important as
nonfactorizable contributions from the hard-gluon exchange, if not even the 
dominant ones. Soft nonfactorizable 
contributions  are of the positive sign, and the same seems to be valid also for the 
hard corrections. 
While hard nonfactorizable corrections have an imaginary part, in the calculation of 
soft contributions the penguin topologies as potential sources  
for the appearance of an imaginary phase were not discussed, but they are expected to be small.

\section{\label{sec:conc}Conclusion}

We have discussed the nonfactorizable contributions to the $B \rightarrow J/\psi K$ decay and 
have calculated leading soft nonfactorizable corrections using QCD light-cone sum rules.  
In spite of theoretical uncertainties involved by application of LCSR method to the 
$B \rightarrow J/\psi K$ decay discussed in Section~\ref{sec:calc}, 
and a possible influence higher 
charmonium resonances to the sum rule, 
the predicted correction clearly favors the positive 
value for $\tilde{F}_{BK}^+$ and therefore of $a_2$.  
%Therefore, our result is in agreement with all recent 
%experimental and theoretical consideration of $a_2$ parameter. 

Recent first observations of the color-suppressed decays of the type 
$\overline{B}^0 \rightarrow D^{(*)0} \pi^0$ by CLEO \cite{CLEO} 
and BELLE \cite{BELLE} indicate also the positive value for $a_2$ parameter. 
%the large final state interaction phases in these decays. 
Although these data show that $a_2$ is a process dependent quantity,
 which is clearly exhibited by the difference in 
the prediction for $a_2$  in $\overline{B}^0 \rightarrow D^{(*)0} \pi^0$ and 
$B \rightarrow J/\psi K$ decays by almost a factor 2 
($|a_2(\overline{B}^0 \rightarrow D^{(0(*))} \pi^0)| = 0.57 \pm 0.06$ {\it vs} 
$|a_2(B \rightarrow J/\psi K)| =  0.28 \pm 0.03$), the positive value for 
$a_2$ can be clearly deduced in both cases. 
%recent theoretical considerations and 
%experimental data on $B$ decays seem to favor positive value for $a_2$ in color-suppressed $B$ decays. 
This is just opposite to the predicted negative values of this parameter in $D$ meson decays. The tendency 
to a positive value of $a_2$ in $B$ decays  
was also observed in the global fit of decay amplitudes to the data 
\cite{NS}, where the arguments in favor of a sign 
change of $a_2$ from negative to the positive when going from $D$ to $B$ decays were presented. 

Moreover, these recent experimental results on $\overline{B}^0 \rightarrow D^{(\ast)0}\pi^0$ point 
out large nonfactorizable contributions, as well as the large final state 
interaction phases in the color-suppressed (class-II) decays \cite{NP}. 
Soft corrections obtained in this paper add up to this picture, being significantly larger 
than soft corrections in the $B \rightarrow \pi \pi $ decay. 

%Inspite of the fact that the predicted value for $a_2$, (\ref{eq:a2calc}) calculated 
%from light-cone sum rules or (\ref{eq:a2final}) corrected by inclusion of 
%hard twist-2 corrections from \cite{Cheng} is still far from the experimentally favored value 
%(\ref{eq:a2exp}), it provides us a clear signal 
%for the large nonfactorizable contributions in the $B \rightarrow J/\psi K$ decay. 

\section*{Acknowledgment}

I would like to thank A. Khodjamirian and R. R{\"u}ckl for numerous fruitful 
discussions and comments. The support by the Alexander 
von Humboldt Foundation and partially support of the Ministry of Science and Technology of the Republic of 
Croatia under the contract 0098002 is greatfully acknowledged.

\end{document}